\documentclass[prb,aps,twocolumn,showpacs,ams]{revtex4}
\usepackage{graphicx}
\usepackage{dcolumn}
\usepackage{latexsym}
\usepackage{color}

\begin{document}


\title{
Transverse magnetic field and chiral-nonchiral transition in vortex states 
for nearly $B \parallel ab$ in chiral $p$-wave superconductors
}



\author{Masahiro Ishihara} 
\affiliation{
Department of Physics, Okayama University, Okayama 700-8530, JAPAN}

\author{Yuujirou Amano} 
\affiliation{
Department of Physics, Okayama University, Okayama 700-8530, JAPAN}

\author{Masanori Ichioka}
\email[]{ichioka@cc.okayama-u.ac.jp}
\affiliation{
Department of Physics, Okayama University, Okayama 700-8530, JAPAN}

\author{Kazushige Machida} 
\affiliation{
Department of Physics, Okayama University, Okayama 700-8530, JAPAN}


\date{\today}

\begin{abstract}
On the basis of Eilenberger theory, 
we study the vortex state when a magnetic field is applied 
nearly parallel to the $ab$ plane in a chiral $p$-wave superconductor 
with a large anisotropy ratio of $ab$ and $c$, as in ${\rm Sr_2RuO_4}$. 
We quantitatively estimate the field dependence of the pair potential, 
magnetization, and flux line lattice form factor, 
and study the transition from the chiral $p_-$ state at low fields 
to the nonchiral $p_y$ state at high fields. 
Even for exactly parallel fields to the $ab$ plane, 
transverse fields exist in the chiral state. 
The chiral-nonchiral transition disappears 
when the magnetic field orientation is 
tilted within $1^\circ$ from the $ab$ plane. 
This may be a reason why the experimental detection of 
this transition is difficult.
\end{abstract}

\pacs{74.25.Uv, 74.70.Pq, 74.20.Rp, 74.25.Ha}


\maketitle

\section{Introduction}
\label{sec:Introduction}

Chiral $p$-wave superconductivity attracts much attention 
as one of representative topological superconductors.~\cite{Nayak} 
The chiral $p$-wave superconductivity with the pairing 
function $p_\pm (=p_x \pm {\rm i}p_y)$ is a possible pairing state, 
when the $p$-wave pairing interaction works instead of 
the conventional $s$-wave pairing. 
The pairing function $p_\pm$ 
breaks time-reversal symmetry, inducing spontaneous magnetic fields 
observed by ${\rm \mu SR}$ experiments.~\cite{Luke} 
We also expect that Majorana states are accommodated at vortices and surfaces 
in chiral $p$-wave superconductors.~\cite{ReadGreen,Ivanov}  
This type of pairing is realized in the A phase of superfluid ${\rm ^3He}$, 
and is a candidate for the superconducting phase of 
${\rm Sr_2RuO_4}$.~\cite{MachenzieMaeno,MaenoJPSJ} 
However, there remain mysteries for the pairing symmetry of ${\rm Sr_2RuO_4}$, 
since we have not observed some typical phenomena expected 
in chiral $p$-wave superconductors. 
For example, when a magnetic field $\bar{\bf B}$ is applied in the orientation 
$\bar{\bf B} \parallel ab$, 
theoretically we expect the transition from the chiral $p_\pm$-wave state 
at low fields to the nonchiral $p_y$-wave state 
(when $\bar{\bf B} \parallel y$) 
at high fields.~\cite{Agterberg,Kaur} 
That is, at low fields, the free energy of the chiral $p_\pm$-wave state 
is lower than that of nonchiral $p_x$- or $p_y$-wave states, 
because the latter nonchiral states have vertical line nodes. 
On the other hand, the nonchiral state is realized at high fields, 
because the upper critical field $H_{\rm c2}$ of the $p_y$ state 
is higher than that of 
the chiral $p_\pm$-wave state when $\bar{\bf B} \parallel y$. 
While the chiral-nonchiral transition was suggested 
by experiments of the magnetization curve,~\cite{Tenya} 
this transition was not observed in other experimental 
methods.~\cite{MaenoJPSJ,Deguchi,Yaguchi} 
There were discussions in that 
the double transition~\cite{MachenzieMaeno,MaenoJPSJ,Yaguchi} 
near $H_{\rm c2}$ corresponds to the chiral-nonchiral transition.~\cite{Kaur}

On the other hand, in superconductors with uniaxial anisotropy, 
transverse magnetic fields appear in the vortex state 
when the field orientation is tilted from the $ab$ plane.\cite{Thiemann}  
This transverse field is detected by the spinflip scattering of 
the small angle neutron scattering (SANS) in the vortex states, 
as demonstrated in ${\rm YBa_2Cu_3O_{7-\delta}}$.~\cite{Kealey}  
Recently, the spin flip SANS by the transverse field was reported 
in ${\rm Sr_2RuO_4}$.~\cite{Rastovski}  
Therefore, the quantitative theoretical estimate of 
the transverse field is expected. 
It is also important to find 
new phenomena by the contribution of Cooper pair's angular momentum 
$L_z/\hbar=\pm 1$ of the $p_\pm$-wave paring. 

The purpose of this study is to establish quantitative theoretical 
estimations of the vortex structure in chiral $p$-wave superconductors 
when a magnetic field is applied to exactly $\bar{\bf B} \parallel ab$, 
and when the field orientation is slightly tilted from the $ab$ plane. 
On the basis of Eilenberger theory by which we can quantitatively calculate 
the spatial structure and the physical quantities of the vortex 
states,~\cite{Hiragi,Klein,Miranovic,IchiokaP}      
we will clarify behaviors of the chiral-nonchiral transition and 
the transverse field structure as a function of a magnetic field $\bar{B}$. 

\section{Formulation by Eilenberger theory}
\label{sec:formulation}

As a model of the Fermi surface, we use a quasi-two dimensional 
Fermi surface with a rippled cylinder shape. 
The Fermi velocity is assumed to be 
${\bf v}=(v_a,v_b,v_c)\propto(\cos\phi,\sin\phi,\tilde{v}_z \sin p_c)$ 
at 
${\bf p}=(p_a,p_b,p_c)\propto(p_{\rm F}\cos\phi, p_{\rm F}\sin\phi,p_c)$
on the Fermi surface.~\cite{Hiragi}  
We consider a case $\tilde{v}_z=1/60$,
producing large anisotropy ratio of coherence lengths, 
$\gamma \equiv \xi_{c} / \xi_{b} \sim 
\langle v_c^2 \rangle_{\bf p}^{1/2}/\langle v_b^2 \rangle_{\bf p}^{1/2}
\sim 1/60$,~\cite{Rastovski}     
where $\langle \cdots \rangle_{\bf p}$ 
indicates an average over the Fermi surface. 
The magnetic field is tilted within $1^\circ$ from the $ab$ plane. 
Since we set the $z$ axis to the vortex line direction, 
the coordinate $(x,y,z)$ for the vortex structure is related to the 
crystal coordinate $(a,b,c)$  
as $(x,y,z)=(a,b \cos\theta + c \sin\theta,c \cos\theta -b \sin\theta)$ 
with $\theta=90^\circ \sim 89^\circ$. 

In a chiral $p$-wave superconductor, 
the pair potential takes the form,
\begin{eqnarray}
\Delta({\bf p},{\bf r})
=\Delta_+({\bf r}) \phi_+({\bf p}) +\Delta_-({\bf r}) \phi_-({\bf p})
\end{eqnarray} 
with the pairing functions 
$\phi_\pm({\bf p})=(p_a \pm {\rm i}p_b)/p_{\rm F}={\rm e}^{\pm{\rm i}\phi} $.  
$\Delta_\pm({\bf r})$ describes the vortex structure 
as a function of ${\bf r}$ (the center of mass coordinate of the pair).
In our study, $\Delta_-({\bf r})$ is a main component and 
$\Delta_+({\bf r})$ is a passive component induced around 
a vortex.~\cite{IchiokaP,IchiokaGL}   
At a zero field, $\Delta_+({\bf r})=0$.
When we consider the $p_x$- and $p_y$-wave components, the pair potential is 
decomposed as 
$\Delta({\bf p},{\bf r})
=\Delta_x({\bf r}) \phi_x({\bf p}) +\Delta_y({\bf r}) \phi_y({\bf p})$ 
with $\phi_x({\bf p})=\sqrt{2} p_a =\sqrt{2}\cos\phi $ and   
$\phi_y({\bf p})=\sqrt{2} p_b =\sqrt{2}\sin\phi $. 

\begin{figure}
\begin{center}
\includegraphics[width=5.5cm]{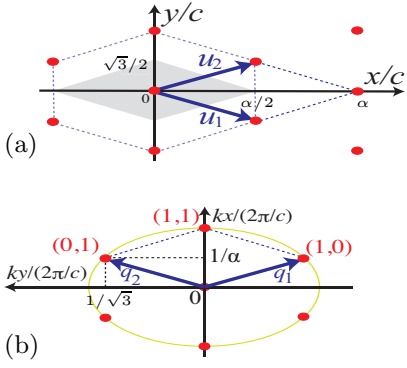}
\end{center}
\caption{\label{fig1}
(Color online) 
(a) Unit vectors ${\bf u}_1$ and ${\bf u}_2$ of the vortex lattice. 
Circles indicate the vortex centers. 
The gray region is a unit cell of our calculations. 
(b) Unit vectors ${\bf q}_1$ and ${\bf q}_2$ in the reciprocal space. 
Circles indicate the SANS spots $(h,k)$. 
}
\end{figure}

Using the anisotropic ratio 
$\Gamma_\theta \equiv
\xi_{y} / \xi_{x} \sim 
\langle v_y^2 \rangle_{\bf p}^{1/2}/\langle v_x^2 \rangle_{\bf p}^{1/2}
\sim (\cos^2\theta+\gamma^{-2}\sin^2\theta)^{-\frac{1}{2}}$, 
we set the unit vectors of the vortex lattice as 
${\bf u}_1=c({\alpha}/{2},-{\sqrt{3}}/{2})$ and 
${\bf u}_2=c({\alpha}/{2}, {\sqrt{3}}/{2})$ 
with $c^2=2 \phi_0/ (\sqrt{3} \alpha \bar{B})$ and 
$\alpha=3 \Gamma_\theta$,~\cite{Hiragi} 
as shown in Fig. \ref{fig1}(a). 
$\phi_0$ is the flux quantum, and $\bar{B}$ is the flux density.  
As shown in Fig. \ref{fig1}(b), 
the unit vectors in the reciprocal space 
are given by ${\bf q}_1=(2\pi/c)(1/\alpha,-1/\sqrt{3})$ 
and ${\bf q}_2=(2\pi/c)(1/\alpha,1/\sqrt{3})$, 
where spots of the SANS appear. 

Quasiclassical Green's functions 
$f(\omega_n , {\bf p},{\bf r})$,  
$f^\dagger(\omega_n , {\bf p},{\bf r})$,  
$g(\omega_n , {\bf p},{\bf r})$ in the vortex lattice states  
are obtained by solving the Riccati equation, 
which is derived from 
the Eilenberger equation
\begin{eqnarray} &&
\left\{ \omega_n 
+ \hat{{\bf v}} \cdot\left(\nabla+{\rm i}{\bf A} \right)\right\} f 
=\Delta g\,,
\nonumber \\ && 
\left\{ \omega_n
-\hat{{\bf v}}\cdot\left( \nabla-{\rm i}{\bf A} \right)\right\} f^\dagger
=\Delta^\ast g . \quad
\label{eq:Eil}
\end{eqnarray} 
in the clean limit, 
with a normalization condition $g=(1-ff^\dagger)^{1/2}$   
and the Matsubara frequency 
$\omega_n$.~\cite{Hiragi,Klein,Miranovic,IchiokaP}    
That is, we calculate the spatial structure of $g$ 
without using Pesch's approximation.\cite{Pesch,Kaur} 
The normalized Fermi velocity is $\hat{\bf v}={\bf v}/v_{\rm F}$ with 
$v_{\rm F}=\langle {\bf v}^2 \rangle_{{\bf p}}^{1/2}$.
We have scaled the length, temperature, magnetic field, 
and energies 
in units of $\xi_0$, $T_c$,
$B_0$, and $\pi k_{\rm B} T_{\rm c}$, respectively, 
where $\xi_0=\hbar v_{{\rm F}}/2\pi k_{\rm B} T_{\rm c}$, 
$B_0=\phi_0 /2 \pi \xi_0^2$.
The vector potential 
${\bf A}=\frac{1}{2}\bar{{\bf B}}\times{\bf r}+{\bf a}({\bf r})$
is related to the internal field as 
${\bf B}({\bf r})=\nabla\times {\bf A}
 =(B_x({\bf r}),B_y({\bf r}),B_z({\bf r}))$ 
with $\bar{\bf B}=(0,0,\bar{B})$, 
$B_z({\bf r})=\bar{B}+b_z({\bf r})$ and 
$(B_x,B_y,b_z)=\nabla\times {\bf a}$. 
The spatial averages of $B_x$, $B_y$, and $b_z$ are zero.\cite{Rastovski} 

We calculate $\Delta({\bf p},{\bf r})$ by the gap equation 
\begin{eqnarray}
\Delta_\pm({\bf r})
= \lambda_0\, 2T \sum_{\omega_n >0}^{\omega_{\rm c}}
 \left\langle \phi_\pm^\ast({\bf p})  f  \right\rangle_{\bf p} , 
\label{eq:scD}
\end{eqnarray}
where 
$\lambda_0=N_0g_0$ is the dimensionless 
$p$-wave pairing interaction 
in the low-energy band $|\omega_n|\le\omega_{c}$, 
defined by the cutoff energy $\omega_{\rm c}$ as 
$1/\lambda_0 = \ln T+2\,T\sum_{\omega_n>0}^{\omega_{\rm c}}\,\omega_n^{-1}$.
We carry out calculations using the cutoff $\omega_{\rm c}=20 k_{\rm B}T_{\rm c}$. 
The current equation to obtain ${\bf A}$ is given by 
\begin{eqnarray}
{\bf j}({\bf r})  
=\nabla\times(\nabla \times {\bf A})= -\kappa^{-2} \ 2T 
\sum_{\omega_n > 0}\left\langle \hat{{\bf v}} \,{\rm Im}\,g\right\rangle_{\bf p}. 
\label{eq:scH}
\end{eqnarray}
The Ginzburg-Landau (GL) parameter $\kappa$ 
is the ratio of the penetration depth to coherence length for 
$\bar{\bf B}\parallel c$, 
and set to be $\kappa=2.7$ appropriate to 
${\rm Sr_2RuO_4}$.~\cite{MachenzieMaeno}
The case of effective GL parameter 
$\kappa_\theta \sim \kappa \Gamma_\theta$ for a field orientation $\theta$
is reproduced by the anisotropy of $\hat{\bf v}$ in Eq. (\ref{eq:scH}).  
Iterating calculations of Eqs. (\ref{eq:Eil})-(\ref{eq:scH})　
at $T=0.5T_{\rm c}$, 
we obtain self-consistent solutions of 
$\Delta_\pm({\bf r})$, ${\bf A}({\bf r})$, and 
quasiclassical Green's functions.

\section{Exactly parallel field to the basal plane}
\label{sec:H90}

\begin{figure}
\begin{center}
\hspace{-0.5cm}
\includegraphics[width=9.0cm]{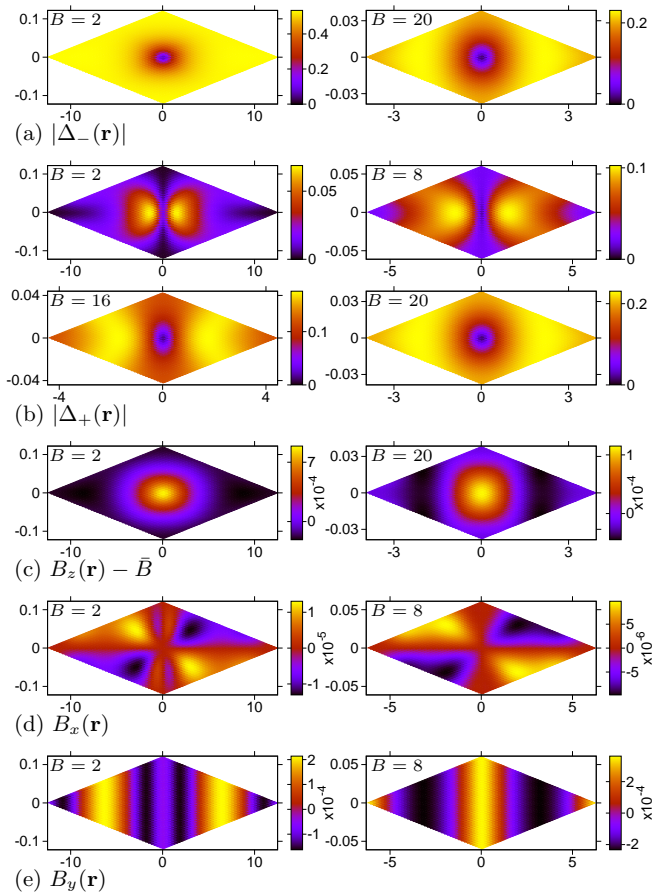}
\end{center}
\caption{\label{fig2}
(Color online) 
Density plots of the pair potential and the internal magnetic field 
within a unit cell [gray region in Fig. \ref{fig1}(a)]  
when $\theta=90^\circ$. 
(a) Main $p_-$ component of the pair potential,  $|\Delta_-({\bf r})|$, 
at $\bar{B}=2$ and 20. 
(b) Passive $p_+$-wave component $|\Delta_+({\bf r})|$ 
at $\bar{B}=2$, 8, 16, and 20. 
(c) $B_z({\bf r})-\bar{B}$ at $\bar{B}=2$ and 20. 
(d) $B_x({\bf r})$ at $\bar{B}=2$ and 8. 
(e) $B_y({\bf r})$ at $\bar{B}=2$ and 8. 
}
\end{figure}
\begin{figure}
\begin{center}
\includegraphics[width=7.0cm]{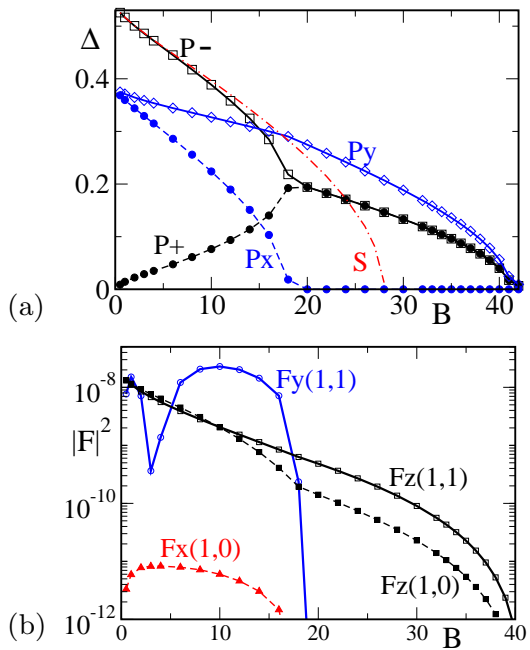}
\end{center}
\caption{\label{fig3}
(Color online) 
(a) $\bar{B}$-dependence of $p_-$, $p_+$, $p_x$, and $p_y$ wave 
components of the pair potential when $\theta=90^\circ$. 
Spatial averaged values of 
$|\Delta_-|$, $|\Delta_+|$, $|\Delta_x|$, and $|\Delta_y|$ are presented. 
We also show the $s$-wave case for comparison. 
(b) $\bar{B}$-dependence of the FLL form factors 
when $\theta=90^\circ$. 
We present $|F_{z}(1,0)|^2$, $|F_{z}(1,1)|^2$, $|F_{x}(1,0)|^2$, 
and $|F_{y}(1,1)|^2$.
Other components $|F_{x}(1,1)|^2$ and $|F_{y}(1,0)|^2$ are less than $10^{-12}$. 
The vertical axis is log-scale. 
}
\end{figure}

First, we study the vortex states 
for exactly $\bar{\bf B} \parallel ab$ ($\theta=90^\circ$). 
In Fig. \ref{fig2}, 
we show the calculated spatial structures 
within a unit cell of the vortex lattice at low and high fields. 
The main component $\Delta_-({\bf r})$ has a winding 1 of the phase 
at the vortex center, 
where the amplitude $|\Delta_-({\bf r})|$ in Fig. \ref{fig2}(a) is suppressed.  
At low fields, the vortex core is localized at the center. 
At high fields the vortex core contribution becomes important 
in the properties of the vortex states, 
since the inter-vortex distances become shorter with increasing fields.  
As a property of chiral $p$-wave superconductors, 
the opposite chiral component $\Delta_+({\bf r})$ 
also appears where the main chiral component $\Delta_-({\bf r})$ has 
spatial modulations around vortex cores.~\cite{IchiokaP,IchiokaGL}    
The amplitude of the induced component $|\Delta_+({\bf r})|$ 
is presented in Fig. \ref{fig2}(b). 
It appears locally around the vortex core at a low field $\bar{B}=2$. 
With increasing fields, 
since the inter-vortex distances become shorter, 
$\Delta_+({\bf r})$ of neighbor vortex cores overlap with each other, 
as shown in panels for $\bar{B}=8$ and 16. 
With further increasing $\bar{\bf B}$, 
the amplitude of $\Delta_+({\bf r})$ is reduced to 
$|\Delta_+({\bf r})|=|\Delta_-({\bf r})|$, 
as shown in a panel for $\bar{B}=20$ in Fig. \ref{fig2}(b). 
This indicates disappearance of $\Delta_x({\bf r})$ by 
the chiral-nonchiral transition from the chiral $p_-$-wave state 
to the nonchiral $p_y$-wave state. 

The $z$-component of the internal field, $B_z({\bf r})$, 
has a conventional spatial structure of 
the vortex lattice also for $\bar{\bf B} \parallel ab$ 
in chiral $p$-wave superconductors, 
if the length is re-scaled by the effective coherence length in each direction. 
As shown in Fig. \ref{fig2}(c), $B_z({\bf r})$ has a peak at a vortex center, 
and decreases as a function of radius from the center. 
We note that the transverse components $B_x({\bf r})$ and $B_y({\bf r})$ 
appear even when exactly $\bar{\bf B} \parallel ab$ 
in the chiral $p_-$ state at low fields. 
This is unconventional behavior 
due to the contribution of the internal angular momentum $L_z$ of 
the chiral pairing function. 
The transverse components vanish in nonchiral $p_y$ states at high fields. 

To see the behaviors of the chiral-nonchiral transition,  
we plot the amplitudes of each component of the pair potential 
as a function of $\bar{B}$ in Fig. \ref{fig3}(a). 
With increasing $\bar{B}$, the $p_-$ wave component decreases 
and the $p_+$ wave component increases. 
After the chiral-nonchiral transition at $B>B^\ast \sim 18$, 
the amplitudes of $p_+$ and $p_-$ are the same.  
If we see the pair potential in the decomposition of $p_x$ and $p_y$, 
with increasing $\bar{B}$, 
the $p_y$ component decreases toward zero at $H_{\rm c2}$  
and the $p_x$ component decreases toward zero at $B^\ast$. 
$\Delta_x=0$ at $\bar{B}>B^\ast$ by the chiral-nonchiral transition. 
In Fig. \ref{fig3}(a), 
we also show the case of conventional $s$-wave pairing. 
Compared with the $s$-wave case, $H_{\rm c2}$ is enhanced 
in the $p_y$-wave state. 
This comes from the fact that $H_{\rm c2}$ is enhanced 
when the pairing function has a horizontal line node 
relative to the field direction.~\cite{Scharnberg}  

To discuss the $\bar{B}$-dependence of the internal field distribution, 
we consider flux line lattice (FLL) form factors  
${\bf F}({\bf q}_{h,k})=(F_{x}(h,k),F_{y}(h,k),F_{z}(h,k))$ 
calculated as Fourier transformation  
of the internal field distribution,  
${\bf B}({\bf r})=\sum_{h,k}{\bf F}({\bf q}_{h,k}) 
\exp({\rm i}{\bf q}_{h,k}\cdot{\bf r})$  
with the wave vectors ${\bf q}_{h,k}=h{\bf q}_1+k{\bf q}_2$.
$h$ and $k$ are integers. 
The $z$-component $|F_{z}(h,k)|^2$ from $B_z({\bf r})$ 
gives the intensity of spots in the conventional non-spinflip 
SANS experiments.\cite{Riseman}  
The transverse component, $|F_{\rm tr}(h,k)|^2=|F_{x}(h,k)|^2+|F_{y}(h,k)|^2$,  
is accessible by the spin-flip SANS experiments.\cite{Kealey,Rastovski} 
In Fig. \ref{fig3}(b), 
we see exponential decays of $|F_{z}(h,k)|^2$ as a function of $\bar{B}$, 
as in the conventional behavior of the vortex states, 
since we do not take care of the Pauli-paramagnetic effect.~\cite{IchiokaPara} 
The transverse components $|F_{x}(1,0)|^2$ and $|F_{y}(1,1)|^2$ 
appear only in the chiral states at $\bar{B}<B^\ast$. 
From the stripe pattern in the spatial structure of $B_y({\bf r})$ 
as in Fig. \ref{fig2}(e), the main spot of $|F_y|^2$ is $|F_{y}(1,1)|^2$. 
The stripe pattern changes between $\bar{B}=2$ and 8. 
From the spatial pattern of $B_x({\bf r})$, 
the main spot of $|F_{x}(h,k)|^2$ is at $(h,k)=(1,0)$. 
The intensity of $|F_{x}(1,0)|^2$ is much smaller than $|F_{y}(11)|^2$. 

\section{Field orientation tilted from the basal plane}
\label{sec:H89}

\begin{figure}
\begin{center}
\hspace{-0.5cm}
\includegraphics[width=9.0cm]{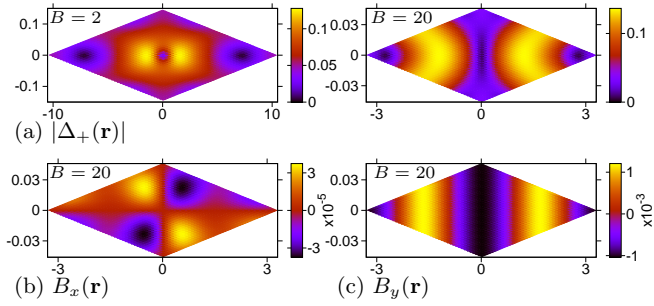}
\end{center}
\caption{
\label{fig4}
(Color online) 
Density plots of the pair potential and the internal magnetic field 
within a unit cell 
when 
$\theta=89^\circ$. 
(a) $|\Delta_+({\bf r})|$ at $\bar{B}=2$ and 20. 
(b) $B_x({\bf r})$ at $\bar{B}=20$. 
(c) $B_y({\bf r})$ at $\bar{B}=20$. 
}
\end{figure}
\begin{figure}
\begin{center}
\includegraphics[width=6.5cm]{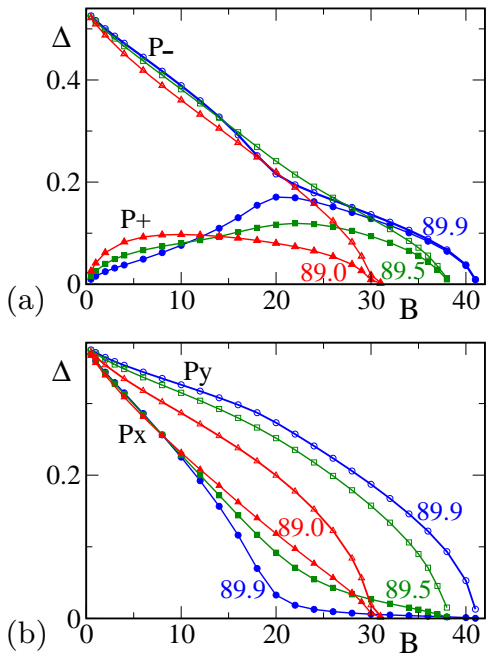}
\end{center}
\caption{\label{fig5}
(Color online) 
$\bar{B}$-dependence of each component of 
the pair potential in the vortex states, 
when the magnetic field is slightly tilted from the $ab$ plane, i.e., 
$\theta=89.9^\circ$, $89.5^\circ$, and $89.0^\circ$. 
(a) Spatial averaged $|\Delta_-|$ and $|\Delta_+|$.  
(b) Spatial averaged $|\Delta_x|$ and $|\Delta_y|$.  
}
\end{figure}
\begin{figure}
\begin{center}
\includegraphics[width=6.0cm]{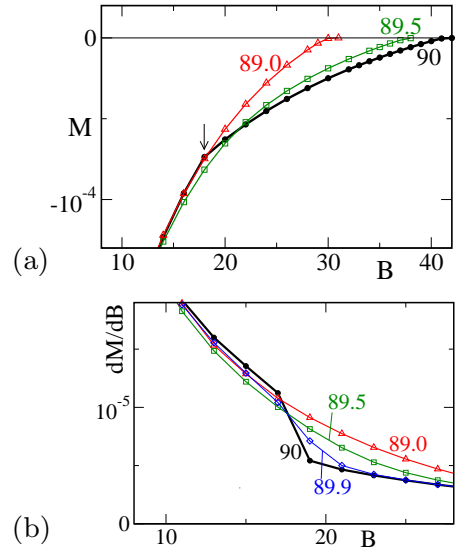}
\end{center}
\caption{\label{fig6}
(Color online) 
(a) Magnetization curve $M(\bar{B})$ as a function of $\bar{B}$ 
when the magnetic field is slightly tilted from the $ab$ plane  
as $89^\circ \le \theta \le 90^\circ$. 
An arrow indicates $B^\ast$. 
(b) Derivative of the magnetization curve ${\rm d}M/{\rm d}\bar{B}$ 
near $B^\ast$ for $89^\circ \le \theta \le 90^\circ$. 
}
\end{figure}
\begin{figure}
\begin{center}
\includegraphics[width=6.0cm]{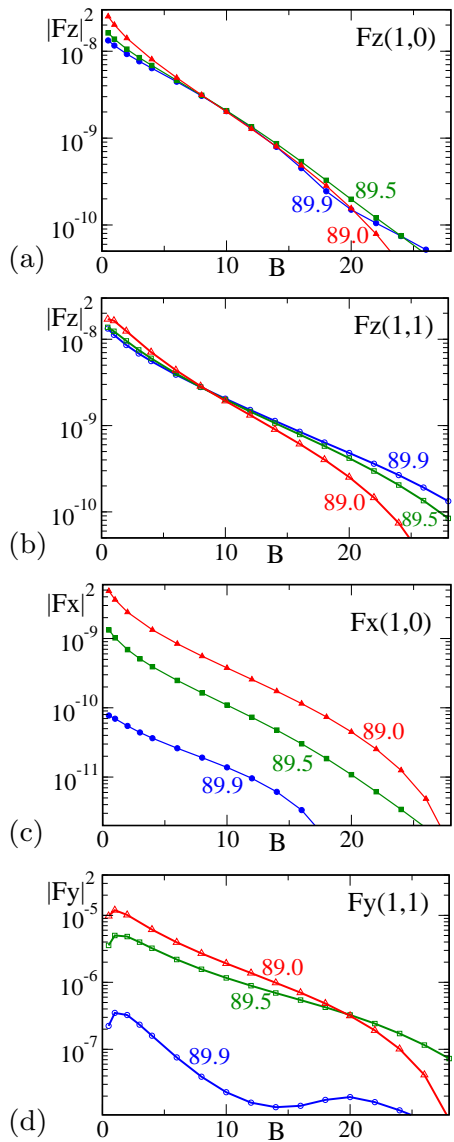}
\end{center}
\vspace{-0.7cm}
\caption{\label{fig7}
(Color online) 
$\bar{B}$-dependence of the FLL form factors in the vortex states, 
when the magnetic field is slightly tilted from the $ab$ plane, i.e., 
$\theta=89.9^\circ$, $89.5^\circ$, and $89.0^\circ$. 
(a) $|F_{z}(1,0)|^2$. 
(b) $|F_{z}(1,1)|^2$. 
(c) $|F_{x}(1,0)|^2$. 
(d) $|F_{y}(1,1)|^2$. 
In (a)-(d), the vertical axis is log-scale. 
}
\end{figure}

Next, we discuss the vortex states when the magnetic field is 
slightly tilted from the $ab$ plane as $89^\circ \le \theta <90^\circ$. 
The vortex states in the $p_-$-wave domain,  
where $\Delta_-({\bf r})$ is the main component,  
has lower free energy than that in the $p_+$-wave domain where 
$\Delta_+({\bf r})$ is the main component. 
This is because the field orientation lifts up the degeneracy 
of the $p_-$- and $p_+$-wave domains.\cite{IchiokaP,IchiokaGL}   
Therefore, we study the stable $p_-$-wave domain case here.  

When $\theta=89^\circ$, around the vortex core of the main component 
$\Delta_-({\bf r})$, the opposite chiral 
component $\Delta_+({\bf r})$ is also induced as presented 
in Fig. \ref{fig4}(a). 
There, the spatial pattern of $|\Delta_-({\bf r})|$ at $\bar{B}=2$ 
is similar to that of $\bar{\bf B}\parallel c$ case,~\cite{IchiokaP}  
rather than that of $\bar{\bf B}\parallel ab$ in Fig. \ref{fig2}(b). 
$|\Delta_+({\bf r})|$ at a high field $\bar{B}=20$ keeps similar 
spatial structure to that of $\bar{\bf B}\parallel ab$ case 
with $\bar{B}=8$ in Fig. \ref{fig2}(b).  
Thus $|\Delta_+({\bf r})| \ne |\Delta_-({\bf r})|$ even at high fields, 
indicating that the nonchiral state with $\Delta_x({\bf r}) = 0$ 
does not realize.  
To see the disappearance of the chiral-nonchiral transition, 
we study the $\bar{B}$-dependence of each component of the pair potential 
for $\theta=89.9^\circ$, $89.5^\circ$, and $89.0^\circ$. 
As seen from the curve for $89.9^\circ$ in Fig. \ref{fig5}(a), 
even if the field orientation is tilted by $0.1^\circ$, 
the chiral-nonchiral transition changes to a crossover behavior. 
At high fields, small differences between 
the $p_-$ and $p_+$ components still exist.  
Thus, in Fig. \ref{fig5}(b), 
a small amplitude of $p_x$-wave component survives up to $H_{\rm c2}$. 
Further tilting the field orientation to $89.5^\circ$ and $89.0^\circ$, 
the crossover behaviors are smeared, 
and we can not see the remnant of the chiral-nonchiral transition anymore. 
The amplitude of the $p_x$-wave component monotonically decreases 
toward $H_{\rm c2}$. 

The chiral-nonchiral transition 
is reflected by the magnetization curve, 
$M=\bar{B}-H$ as a function of $\bar{B}$. 
From the selfconsistent solutions we obtain the relation of $\bar{B}$ and 
the external field $H$ as 
\begin{eqnarray} && 
H=
\bar{B}
+\left\langle \left( B_z({\bf r})-\bar{B} \right)^2\right\rangle_{\bf r}
/{\bar{B}} 
\nonumber \\ &&   
+\frac{T}{{\kappa}^2 \bar{B}} \sum_{\omega_n >0} \langle \langle  
{\rm Re} \{ 
\frac{(f^\dagger \Delta+f \Delta^\ast)g}{2(g+1)} 
+\omega_n ( g-1 ) \} 
\rangle_{\bf p}\rangle_{\bf r}, \qquad 
\end{eqnarray} 
which is derived by Doria-Gubernatis-Rainer scaling.~\cite{WatanabeKita,Doria} 
$\langle\cdots\rangle_{\bf r}$ indicates a spatial average. 
In the magnetization curve in Fig. \ref{fig6}(a), 
we see a change of the slope at $B^\ast$ 
for exactly $\bar{\bf B} \parallel ab$ ($\theta=90^\circ$).
This is clearly seen as a step at $B^\ast$ 
in the plot of the derivative ${\rm d}M/{\rm d}B$ in Fig. \ref{fig6}(b). 
However this behavior of second-order phase transition is smeared 
by tilting the field orientation within $1^\circ$. 
The step in ${\rm d}M/{\rm d}B$ was suggested in 
the experimental observation of the magnetization curve.~\cite{Tenya} 
However in other experimental methods such as specific 
heat and thermal conductivity,~\cite{Deguchi,MaenoJPSJ,Yaguchi} 
the chiral-nonchiral transition has not been observed yet. 
This may be because the experimental situation of exactly 
$\bar{B}\parallel ab$ is difficult to be realized. 
Our study shows that the chiral-nonchiral transition vanishes  
by tilting the field orientation within $1^\circ$. 

In ${\rm Sr_2RuO_4}$, when ${\bar B}\parallel ab$, 
$H_{\rm c2}$ is suppressed and changes to the first order phase 
transition.~\cite{Tenya,Kittaka,Yonezawa} 
Our simple formulation in this work does not include the mechanism 
for the suppression of $H_{\rm c2}$, such as a Pauli-paramagnetic-like 
effect.\cite{Machida} 
The study for this $H_{\rm c2}$ behavior belongs to future works. 

Both in the $\bar{B}$-dependence of $|F_{z}(10)|^2$ 
in Fig. \ref{fig7}(a) and $|F_{z}(11)|^2$ 
in Fig. \ref{fig7}(b), 
with decreasing $\theta$ from $90^\circ$, 
$|F_z|^2$ becomes larger at low fields,  
reflecting the decrease of the effective GL parameter $\kappa_\theta$.  
Roughly $|F_z| \propto (\kappa \kappa_\theta)^{-1}$ from Eq. (\ref{eq:scH}). 
On the other hand, $|F_z|^2$ becomes smaller at high fields, 
because $H_{\rm c2}$ decreases by the decrease of $\theta$. 
As shown in Figs. \ref{fig4}(b)-(c), 
$B_x({\bf r})$ and $B_y({\bf r})$ have similar spatial structures  
until high fields to those of $\bar{B}=2$ and $\theta=90^\circ$ 
in Figs. \ref{fig2}(d)-(e). 
However, the amplitudes of $B_x$ and $B_y$ at $\theta=89^\circ$ 
are much larger than those at $\theta=90^\circ$. 
Thus, $|F_{x}(1,0)|^2$ in Fig. \ref{fig7}(c) and $|F_{y}(1,1)|^2$ 
in Fig. \ref{fig7}(d) increase rapidly with decreasing $\theta$ 
from $90^\circ$. 
$|F_{x}(1,1)|^2$ and $|F_{y}(1,0)|^2$ are less than $10^{-12}$. 

When these form factors are compared with each other, 
the intensity of the spinflip SANS at ${\bf q}_{1,1}$ from $|F_{y}(1,1)|^2$ 
is much larger than that of the non-spinflip SANS intensity 
of $|F_{z}(1,1)|^2$ and $|F_{z}(1,0)|^2$. 
On the other hand, very small intensity of the spinflip SANS at 
${\bf q}_{1,0}$ from $|F_{x}(1,0)|^2$ is difficult to be observed. 
These correspond to the SANS experimental results on ${\rm Sr_2RuO_4}$, 
where the spinflip SANS spot was observed only 
at ${\bf q}_{1,1}$.\cite{Rastovski} 
Within the experimental resolution, the spin-flip SANS spot at ${\bf q}_{1,0}$ 
and the non-spinflip SANS spots have not been observed yet.  
We note that similar behaviors of the transverse fields appear 
also in the nonchiral state including $s$-wave pairing,  
if $\theta \ne 90^\circ$. 
Thus, for $\theta \ne 90^\circ$, 
it is not easy that unique effects due to the chiral state are extracted 
from qualitative behaviors of the transverse fields 

\section{Summary}
\label{sec:summary}

We studied the vortex states for nearly $\bar{\bf B}\parallel ab$ 
in chiral $p$-wave superconductors on the basis of Eilenberger theory. 
The chiral-nonchiral transition at exactly $\bar{\bf B}\parallel ab$ 
vanishes by tilting the magnetic field within $1^\circ$. 
We quantitatively estimated the FLL form factors including transverse fields,  
and showed that 
the spin-flip SANS intensity by the transverse fields 
has large intensity at $(1,1)$-spot. 
The transverse fields appear even when exactly $\bar{\bf B}\parallel ab$, 
as a unique effect of the chiral states.  
These theoretical results indicate the importance of careful studies  
about the vortex states for nearly $\bar{\bf B} \parallel ab$, 
to detect some natures of chiral $p$-wave superconductors or ${\rm Sr_2RuO_4}$.

\begin{acknowledgments}
We thank M. R. Eskildsen for fruitful discussions and 
information about their spin-flip SANS experiments.
\end{acknowledgments}


\end{document}